\documentclass[10pt, conference, compsocconf]{IEEEtran}
\usepackage{algorithm,algorithmic,amsbsy,amsmath,amssymb,epsfig,bbm,mathrsfs,mathdots,subfigure,arydshln,arydshln,multirow,verbatim,booktabs,setspace,url,stfloats,fixltx2e,cite,nomencl,epstopdf}
\usepackage{xcolor}
\def\BibTeX{{\rm B\kern-.05em{\sc i\kern-.025em b}\kern-.08em
    T\kern-.1667em\lower.7ex\hbox{E}\kern-.125emX}}

\newcommand{\beq}{\begin{equation}}
\newcommand{\eeq}{\end{equation}}
\newcommand{\bitm}{\begin{itemize}}
\newcommand{\ba}{\begin{array}}
\newcommand{\ea}{\end{array}}
\newcommand{\eitm}{\end{itemize}}
\newcommand{\beqn}{\begin{eqnarray}}
\newcommand{\eeqn}{\end{eqnarray}}
\newcommand{\beqno}{\begin{eqnarray*}}
\newcommand{\eeqno}{\end{eqnarray*}}
\newcommand{\bma}{\begin{displaymath}}
\newcommand{\ema}{\end{displaymath}}
\newcommand{\bnu}{\begin{enumerate}}
\newcommand{\enu}{\end{enumerate}}
\newcommand{\bce}{\begin{center}}
\newcommand{\ece}{\end{center}}
\newcommand{\btb}{\begin{tabular}}
\newcommand{\etb}{\end{tabular}}

\begin{document}
\title{Adaptive Resource Allocation in Quantum Key Distribution (QKD) for Federated Learning}
\author{\IEEEauthorblockN{Rakpong Kaewpuang$^{\mathrm{1}}$, Minrui Xu$^{\mathrm{1}}$, Dusit Niyato$^{\mathrm{1}}$, Han Yu$^{\mathrm{1}}$, Zehui Xiong$^{\mathrm{2}}$, and Xuemin (Sherman) Shen$^{\mathrm{3}}$} \\
\IEEEauthorblockA{ $^{\mathrm{1}}$School of Computer Science and Engineering, Nanyang Technological University (NTU) \\
				   $^{\mathrm{2}}$Pillar of Information Systems Technology and Design, Singapore University of Technology and Design (SUTD) \\
                  $^{\mathrm{3}}$Department of Electrical and Computer Engineering, University of Waterloo   
       }}
\maketitle

\begin{abstract}
Increasing privacy and security concerns in intelligence-native 6G networks require quantum key distribution-secured federated learning (QKD-FL), in which data owners connected via quantum channels can train an FL global model collaboratively without exposing their local datasets. To facilitate QKD-FL, the architectural design and routing management framework are essential. However, effective implementation is still lacking. To this end, we propose a hierarchical architecture for QKD-FL systems in which QKD resources (i.e., wavelengths) and routing are jointly optimized for FL applications. In particular, we focus on adaptive QKD resource allocation and routing for FL workers to minimize the deployment cost of QKD nodes under various uncertainties, including security requirements. The experimental results show that the proposed architecture and the resource allocation and routing model can reduce the deployment cost by 7.72\% compared to the CO-QBN algorithm.
\end{abstract}


\section{Introduction}
\label{sec:introduction}

Recent security threats from quantum computers are leading to a revolution in the use of quantum channels for secret key distribution, i.e., quantum key distribution (QKD), for the transmission of confidential information~\cite{cao2022evolution}. In intelligence-native 6G communications~\cite{you2021towards}, security and privacy are crucial requirements in distributed artificial intelligence (AI) training. On the one hand, federated learning (FL) enables multiple FL workers to train a global model while keeping their data local~\cite{lim2020federated,yang2019federated}. On the other hand, QKD distributes secure keys for FL nodes to encrypt the model parameters~\cite{mehic2020quantum}. However, the resource allocation problem in QKD-FL remains a critical challenge that prevents the implementation of such secure communication systems.

For resource allocation in QKD-FL systems, researchers have proposed solutions that must be effective~\cite{huang2021starfl}. For example, in~\cite{xu2021multi}, for the problem of selecting and routing among FL nodes, the authors introduced a resource allocation framework based on federated reinforcement learning, where the model transmissions of remote devices are protected by adaptive jamming. For the QKD resource allocation problem, the shortest path-based heuristic algorithm was proposed for measurement-device-independent-QKD (MDI-QKD) networks~\cite{cao2021hybrid}. However, the above works solve the routing problem and the resource allocation problem separately, which leads to nonlinear and inefficient resource allocation issues in quantum-safe FL systems. Moreover, the current works on resource allocation schemes for quantum Internet mainly focus on developing static algorithms~\cite{xu2022quantum, cao2019cost}, while the uncertainty in quantum states and quantum communication is largely neglected.

To address these issues, in this paper, we propose a hierarchical architecture for QKD-FL systems in which adaptive QKD resources and routing are jointly considered. In this framework, the QKD network manager performs joint optimization of resource allocation and routing policies in the QKD network to provide services for unpredictable security requirements, i.e., dynamic secret-key rates, in the FL network. To handle uncertain factors, we formulate the joint assignment and routing problem as a two-stage stochastic programming model. Therefore, adaptive decision-making in the reservation and on-demand stages can efficiently enable quantum-protected FL applications to obtain sufficient quantum secure keys.

The major contributions of this paper can be summarized as follows:
\begin{itemize}
    \item We introduce a hierarchical architecture for QKD-FL systems, where resource allocation and routing decisions are coupled. To supply enough secret keys for machine learning models during collaborative training among FL nodes, the QKD network managers allocate QKD resources from QKD nodes and determine routing strategies for QKD links.
    \item We formulate and solve the stochastic programming (SP) model to obtain the optimal decisions on resource allocation (i.e., QKD and KM wavelengths) and routing from a limited resource pool. In the proposed model, adaptive QKD resource allocation and routing are performed with prior knowledge in the reservation stage and realization in the on-demand stage.
    \item We highlight the superiority of the proposed model by evaluating the performance under real-world network topology and comparing the solution of the proposed model with an existing baseline algorithm.
\end{itemize}

\section{Related Works}
\label{sec:related-workds}

\subsection{Federated Learning}

As a growing concern for user privacy, FL for wireless edge networks was introduced and attracted the increasing attention of researchers from both physical and application layers~\cite{xu2022edge, you2021towards}. For example, Wang \textit{et al.} in~\cite{wang2020federated} proposed a decentralized cooperative edge caching algorithm based on federated reinforcement learning. This algorithm enables wireless IoT devices to collaboratively train a global model while keeping their private data samples locally. Moreover, Xu \textit{et al.} in~\cite{xu2021multi} developed an intelligent incentive mechanism based on federated reinforcement learning with multiple agents. In this mechanism, the problems of incentivizing FL workers and managing radio resources were jointly addressed by training a global multi-task model collaboratively. However, the training and inference processes of FL is subject to serious security threats, e.g., during the model transfer, from both eavesdropping attacks and the upcoming era of quantum computing, which is able to break traditional encryption methods efficiently.

\subsection{The Quantum Internet}
With the ability to provide information-theoretic security for communications, QKD and QKD-secured systems are expected to protect various 6G communications such as optical, satellite, maritime, and their cross-layer communications~\cite{you2021towards, xu2022quantum}. Unlike traditional QKD protocols, such as Bennett-Brassard-1984 (BB84) and Bennett-Brassard-Mermin-1992 (BBM92)~\cite{cao2022evolution}, modern QKD protocols based on MDI technology can provide longer distribution distance and stronger security without assuming trusted relays~\cite{lo2012measurement}. To minimize the cost of deploying QKD resources in MDI-QKD, Cao~\textit{et al.} in~\cite{cao2021hybrid} proposed a static linear programming model and the CO-QBN algorithm to efficiently manage QKD networks. However, existing QKD resource allocation schemes largely overlook uncertain factors such as user demand for secret key rates, which is unpredictable, especially in FL networks where the FL workers can participate and leave the model training dynamically and spontaneously. Therefore, it is the aim of this paper to optimize the QKD-FL network resources with the objective to minimize total costs while meeting uncertain demand and network conditions. The authors in \cite{kaewpuang2022resource} proposed the resource allocation scheme for quantum-secured space-air-ground integrated networks (SAGIN) in which QKD services protect secure communications between space, aerial, and ground nodes by exchanging secret keys in quantum channels. In \cite{kaewpuang2022resource}, the authors formulated and solved the stochastic programming model to achieve the optimal solution for resource allocation and routing under the uncertainties of the secret-key rate requirements and weather conditions.

\section{System Model}
\label{sec:system-model}
\begin{figure}[htb]
\begin{center}
$\begin{array}{c} \epsfxsize=3 in \epsffile{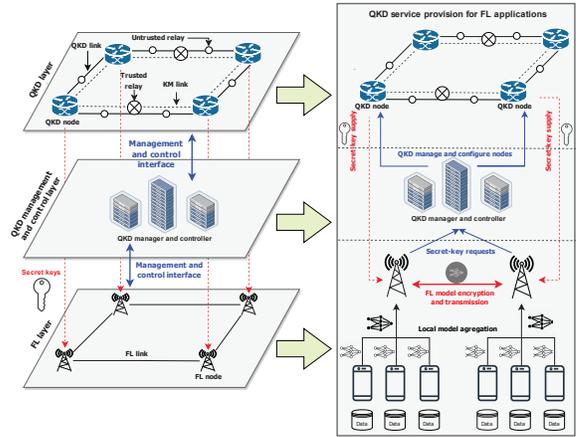} \\
\end{array}$
\caption{The QKD-secured federated learning network architecture.} 
\label{fig:system-model}
\end{center}
\vspace{-0.6cm}
\end{figure}

We propose a three-layer QKD-FL architecture which is illustrated in Fig. \ref{fig:system-model}. The architecture consists of the QKD layer, the QKD management and control layer, and the FL layer. In the QKD layer, there are three types of nodes and two types of links. The three types of nodes are QKD nodes, trusted relays, and untrusted relays. The two types of links are key management (KM) links and QKD links. We assume that QKD nodes are co-located with FL nodes and the links (i.e., QKD, KM, and optical links) are multiplexed within a single fiber by MUX/DEMUX components \cite{cao2021hybrid}. Therefore, the network topology for the QKD layer is the same as that for the FL layer. Let $\mathcal{N}$ and $\mathcal{E}$ denote the set of FL/QKD nodes and the set of fiber links between FL/QKD nodes, respectively.

A QKD node supplies secret keys to its co-located FL node. The trusted relay and the untrusted relay are located between the QKD nodes. The QKD node consists of a global key server (GKS), a local key manager (LKM), a security infrastructure (SI), and components of trusted/untrusted relays. A trusted relay consists of two or multiple transmitters of MDI-QKD (MDI-Txs), LKM, and SI, while an untrusted relay consists of one or multiple receivers of MDI-QKD (MDI-Rxs). An MDI-QRx must be located between two MDI-QTxs to generate local secret keys. An LKM is used to obtain and store the local secret keys from the connected MDI-QTxs, and LKM performs secret key propagation via the one-time-pad (OTP) method \cite{cao2021hybrid} to generate global secret keys between QKD nodes. The SIs are applied to ensure that trusted relays work in a secure manner. In the QKD layer, the QKD links are used to connect between MDI-QTxs and MDI-QRxs, while the KM links are used to connect between LKMs. The QKD link contains quantum and classical channels, while the KM link contains a classical channel implemented by wavelengths. Let $W^{\mathrm{qkd}}_{i,j}$ denote available QKD wavelengths on the QKD link from FL/QKD node $i$ to node $j$ ($i, j \in \mathcal{N}$). Let $W^{\mathrm{kml}}_{i,j}$ denote the available KM wavelengths on the KM link from FL/QKD node $i$ to node $j$.  
     
\subsection{Federated Learning over QKD networks}
\label{subsec:fl-over-qkd}

As shown in Fig. \ref{fig:system-model}, in the FL layer, FL workers engage in FL model training with the assistance of cluster heads \cite{lim2020federated}. FL nodes (e.g., model owners) distribute the initial models to FL workers (e.g., data owners) via the cluster heads (e.g., routers). The initial model is distributed to FL workers. After the FL workers receive the initial model, they train the model locally with their own data and create the new models. Then, the FL workers send their new models to the cluster heads for model aggregation.
 
For the QKD management and control layer, when the FL nodes request secret keys for model encryption from the QKD network manager, QKD nodes controlled by the QKD network controller provide the secret keys for the FL node requests through the QKD network manager. If the QKD nodes cannot provide the secret keys, the FL nodes must wait for the secret keys from the QKD nodes. When the cluster heads receive the local models from the FL workers, they encrypt the aggregated global model with the secret keys and transmit the encrypted model to the destination FL node. Therefore, the security requirements of each FL node in the FL layer directly depend on the available secret key resources in the QKD layer.

\subsection{QKD Network Model for Federated Learning}
\label{subsec:network_model}
In the QKD-FL systems, let $\mathcal{F}$ and $f(s_{f}, d_{f}, P_{f}(\cdot)) \in \mathcal{F}$ denote the set of QKD-FL chain requests and the QKD-FL chain request of FL nodes, respectively. $s_{f}$ and $d_{f}$ are the source node and destination node of QKD-FL chain request $f$, respectively. Let $K_{D}$ denote the maximum achievable secret-key rate at distance $D$. The number of parallel QKD-FL links $P_{f}(\cdot)$ to satisfy the security demand (secret-key rate) of FL model transmission between $s_{f}$ and $d_{f}$ can be expressed as
\begin{equation}
	P_{f}(\tilde{\omega}) = \Big\lceil \frac{\tilde{\kappa}_{f}}{K_{D}}	\Big\rceil, \label{eq:parallel-QKD-FL-links}		
\end{equation}
where $f$ is the QKD-FL chain request and $\tilde{\kappa}_{f}$ is the random variable of secret-key rate requirement of request $f$. $D$ is the distance between two connected MDI-QTxs, which can be expressed as follows:  
\begin{equation}
	D \approx 2 \cdot \vartheta, \label{eq:distance_mdi-qtxs} 			
\end{equation}
where $\vartheta$ is the distance between the MDI-QRx and MDI-QTx. 

\subsection{Cost Model for QKD Network}
\label{subsec:cost_model}

We adopt the costs of the QKD network components and the links from \cite{cao2021hybrid} to support the deployment of QKD with hybrid trusted/untrusted relays on the existing optical backbone network. The QKD network components comprise MDI-QTxs, MDI-QRxs, LKMs, SIs, and multiplexing/demultiplexing (MUX/DEMUX) components. The links comprise QKD links and KM links. The components and links can be described as follows:

\subsubsection{MDI-QTxs and MDI-QRxs} An MDI-QKD process requires two MDI-QTxs and one MDI-QRx and therefore the numbers of MDI-QTxs ($A^{f}_{\mathrm{tx}}(\cdot)$) and MDI-QRxs ($A^{f}_{\mathrm{rx}}(\cdot)$), which satisfy the QKD-FL chain request $f$, can be expressed as follows: 
\beqn
		A^{f}_{\mathrm{tx}} (\tilde{\omega}) = \sum_{i \in \mathcal{N}_{f}} \sum_{j \in \mathcal{N}_{f}} 2 P_{f}(\tilde{\omega}) \Big\lceil \frac{e_{(i,j)}} {D}	\Big\rceil,   \label{eq:cost-mdi-qtxs} \\
		A^{f}_{\mathrm{rx}} (\tilde{\omega}) = \sum_{i \in \mathcal{N}_{f}} \sum_{j \in \mathcal{N}_{f}}   P_{f}(\tilde{\omega}) \Big\lceil \frac{e_{(i,j)}} {D}	\Big\rceil.   \label{eq:cost-mdi-qrxs}
\eeqn
$e_{(i,j)}$ is the distance of the physical fiber link between nodes $i$ and $j$. $\mathcal{N}_{f}$ is set of nodes having the fiber links on the route of request $f$. 

\subsubsection{Local Key Managers (LKMs)} For the QKD-FL chain request $f$, the required number of LKMs ($A^{f}_{\mathrm{km}}$) can be expressed as follows:
\beqn
	   A^{f}_{\mathrm{km}} =  \sum_{i \in \mathcal{N}_{f}} \sum_{j \in \mathcal{N}_{f}} \Big\lceil \frac{e_{(i,j)}} {D} + 1 \Big\rceil.   \label{eq:cost-lkms} 
\eeqn

\subsubsection{Security Infrastructures (SI)} The required number of SIs ($A^{f}_{\mathrm{si}}$) to satisfy an QKD-FL chain request $f$ is expressed follows:
\beqn
	   A^{f}_{\mathrm{si}} =  \sum_{i \in \mathcal{N}_{f}} \sum_{j \in \mathcal{N}_{f}} \Big\lceil \frac{e_{(i,j)}} {D} - 1 \Big\rceil.   \label{eq:cost-si} 
\eeqn
\subsubsection{MUX/DEMUX components} The required number of MUX/DEMUX component pairs ($A^{f}_{\mathrm{md}}$) for the QKD-FL chain request $f$ is expressed as follows:
\beqn
	A^{f}_{\mathrm{md}} =  \sum_{i \in \mathcal{N}_{f}} \sum_{j \in \mathcal{N}_{f}} \Big\lceil \frac{e_{(i,j)}} {D} \Big\rceil  +   \sum_{i \in \mathcal{N}_{f}} \sum_{j \in \mathcal{N}_{f}} \Big\lceil \frac{e_{(i,j)}} {D} - 1 \Big\rceil.  \label{eq:cost-mux-demux}
\eeqn

\subsubsection{QKD and KM links} Three wavelengths and one wavelength are occupied by a QKD link and a KM link, respectively \cite{cao2021hybrid}. The link cost for the QKD-FL chain request $f$ can be expressed as follows: 
\beqn
	A^{f}_{\mathrm{ch}} ( \tilde{\omega} ) =  \sum_{i \in \mathcal{N}_{f}} \sum_{j \in \mathcal{N}_{f}} ( 3 P_{f}(\tilde{\omega}) e_{(i,j)} + e_{(i,j)} ).  \label{eq:cost-qkd-km-link} 
\eeqn
Here, the physical lengths of QKD links and KM links are denoted by $3P_{f}(\tilde{\omega}) e_{(i,j)}$ and $e_{(i,j)}$, respectively.

\section{Optimization Formulation}
\label{sec:optimization-formulation}

\subsection{Model Description}
\label{subsec:model-description}

The optimization is based on the two-stage SP \cite{Brige1997}. The sets and constants in the optimization are defined as follows: 

\begin{itemize}
	\item ${\mathcal{N}}$ denotes a set of all nodes in the network.
	\item ${\mathcal{O}}_n$ denotes a set of outgoing links from node $n \in {\mathcal{N}}$.
	\item ${\mathcal{I}}_n$ denotes a set of incoming links to node $n \in {\mathcal{N}}$.
	\item ${\mathcal{F}}$ denotes a set of QKD-FL chain requests in the network.
	\item $W^{\mathrm{qkd}}_{i,j}$ denotes available QKD wavelengths.
	\item $W^{\mathrm{kml}}_{i,j}$ denotes available KM wavelengths.
	\item $W^{\mathrm{qkd}}_{f}$ denotes 3 wavelengths of QKD link of $f \in {\mathcal{F}}$ satisfying the secrete-key rate.
	\item $W^{\mathrm{kml}}_{f}$ denotes 1 wavelength of KM link of $f \in {\mathcal{F}}$ satisfying the secrete-key rate.
	\item $B^{\mathrm{eng}}_{n, f}$ denotes amount of energy required for transferring traffic of request $f \in {\mathcal{F}}$ going through node $n \in {\mathcal{N}}$.
\end{itemize}
We denote $S_f \in {\mathcal{N}}$ and $D_f \in {\mathcal{N}}$ as the source node and the destination node of request $f$, respectively. 

The decision variables of the network are as follows:

\begin{itemize}
	\item $x_{i,j,f}$ is a binary variable indicating whether request $f \in {\mathcal{F}}$ will take a route with the link from node $i \in {\mathcal{N}}$ to node $j \in {\mathcal{N}}$ or not.
	\item $y^{\mathrm{r}}_{i,n,f}$ and $z^{\mathrm{r}}_{i,n,f}$  are non-negative variables indicating the wavelength channels for QKD and KM links in the reservation phases, respectively.
	\item  $y^{\mathrm{e}}_{i,n,f}$ and $z^{\mathrm{e}}_{i,n,f}$ are non-negative variables indicating the utilized wavelength channels for QKD and KM links, respectively.
	\item  $y^{\mathrm{o}}_{i,n,f}$ and $z^{\mathrm{o}}_{i,n,f}$ are non-negative variables indicating the wavelength channels for QKD and KM links in the on-demand phases, respectively. 
\end{itemize}

We assume that the secret-key rate requirement of request $f$ is uncertain, which is considered to be a random variable. The random variable $\tilde{\omega}$ can be represented by a scenario. Let $\omega$ denote a scenario of request $f$. This scenario is a realization of a random variable $\tilde{\omega}$. The value of the random variable can be taken from a set of scenarios. Let $\Psi$ and $\Omega_{f}$ denote the set of all scenarios of each requirement (i.e., a scenario space) shown in (\ref{eq:scenario_space1}) and the set of all scenarios of request $f$ shown in (\ref{eq:scenario_space2}), respectively. Let $K$ denote the maximum required secret-key rate of the request $f$.
\beqn
 &&\Psi = {\displaystyle \prod_{f \in \mathcal{F}}} \Omega_{f} = \Omega_{1} \times \Omega_{2} \times \cdots \times \Omega_{|\mathcal{F}|}  	\label{eq:scenario_space1}	\\
 & & \Omega_f  = \{ 0, 1, 2,\dots, K \} 	\label{eq:scenario_space2}
\eeqn

The expectation of the SP \cite{Brige1997} can be represented by the weighted sum of scenarios and their probabilities $\mathbb{P}_{f}(\omega)$.

\subsection{Optimization Formulation}
\label{subsec:deterministic-evuivalent}

The SP with the random variable $\tilde{\omega}$ can be transformed into the deterministic equivalence problem \cite{Brige1997} as expressed in (\ref{eq:sp_obj}) - (\ref{eq:def_rt_const14}). 

The objective function (\ref{eq:sp_obj}) is to minimize the deployment cost for QKD resources, consists of MDI-QTxs, MDI-QRxs, LKMs, SIs, MUX /DEMUX components, and QKD and KM links. The decision variables $y^{\mathrm{e}}_{i, n, f, \omega }$, $y^{\mathrm{o}}_{i, n, f, \omega }$, $z^{\mathrm{e}}_{i, n, f, \omega }$, and $z^{\mathrm{o}}_{i, n, f, \omega}$ are under scenario $\omega$ (i.e., $\omega \in \Omega_f$) which implies that the values of demands are available when $\omega$ is observed. The constraint (\ref{eq:def_rt_const1}) ensures that the number of outgoing routes is larger than the number of incoming routes when the node is the source node $S_f$ of the QKD-FL chain request $f$. The constraint (\ref{eq:def_rt_const2}) ensures that the number of incoming routes is larger than the number of outgoing routes if the node is the destination node $D_f$ of the QKD-FL chain request $f$. The constraint (\ref{eq:def_rt_const3}) ensures that the number of outgoing routes must be equal to the number of incoming routes if the node is an intermediate node of the QKD-FL chain request $f$. The constraint (\ref{eq:def_rt_const4}) ensures that there is no loop for any QKD-FL chain request, meaning that there is only one outgoing route for the QKD-FL chain request of any node.  

The constraint (\ref{eq:def_rt_const5}) ensures that the QKD wavelengths of all QKD-FL chain requests of any node must not exceed the available QKD wavelengths (i.e., $W^{\mathrm{qkd}}_{i,j}$). The constraint (\ref{eq:def_rt_const6}) ensures that the KM wavelengths of all QKD-FL chain requests from any node must not exceed the available KM wavelengths (i.e., $W^{\mathrm{kml}}_{i,j}$). The constraint (\ref{eq:def_rt_const7}) ensures that the QKD wavelengths utilized are less than or equal to the QKD wavelengths reserved. The constraint (\ref{eq:def_rt_const8}) ensures that the KM wavelengths utilized are less than or equal to the KM wavelengths reserved. The constraint (\ref{eq:def_rt_const9}) ensures that the QKD wavelengths utilized must satisfy the security requirements (i.e., secret-key rates). The constraint (\ref{eq:def_rt_const10}) ensures that the KM wavelengths used must satisfy the requirements. The constraints (\ref{eq:def_rt_const13}) and (\ref{eq:def_rt_const14}) are the binary and integer variables, respectively.
 
\beqn
	\min	& &	  \sum_{f \in {\mathcal{F}}} \sum_{n \in {\mathcal{N}} } \sum_{i \in {\mathcal{I}}_n } \Big( B^{\mathrm{eng}}_{n, f} x_{i,n,f} + \big( \frac{1}{3} ( A^{f}_{\mathrm{tx}}(\bar{\omega}) \beta^{\mathrm{r}}_{\mathrm{tx}}  \nonumber \\
	& & + A^{f}_{\mathrm{rx}}(\bar{\omega})  \beta^{\mathrm{r}}_{\mathrm{rx}} ) \big) y^{\mathrm{r}}_{i,n,f} + ( A^{f}_{\mathrm{km}} \beta^{\mathrm{r}}_{\mathrm{km}} + A^{f}_{\mathrm{si}} \beta^{\mathrm{r}}_{\mathrm{si}}  \nonumber \\
	& & + A^{f}_{\mathrm{md}} \beta^{\mathrm{r}}_{\mathrm{md}} )z^{\mathrm{r}}_{i,n,f} + e_{(i,n)} ( y^{\mathrm{r}}_{i,n,f} + z^{\mathrm{r}}_{i,n,f} ) \beta^{\mathrm{r}}_{\mathrm{ch}} \Big) \nonumber \label{eq:sp_obj} \\
	& & + \sum_{f \in {\mathcal{F}}} \mathbb{P}_{f}(\omega) \sum_{n \in {\mathcal{N}} } \sum_{i \in {\mathcal{I}}_n } \Big(  \big( \frac{1}{3} ( A^{f}_{\mathrm{tx}}(\omega)\beta^{\mathrm{e}}_{\mathrm{tx}} \nonumber \\
	& & + A^{f}_{\mathrm{rx}}(\omega) \beta^{\mathrm{e}}_{\mathrm{rx}} ) \big) y^{\mathrm{e}}_{i,n,f,\omega} + ( A^{f}_{\mathrm{km}}\beta^{\mathrm{e}}_{\mathrm{km}} + A^{f}_{\mathrm{si}}\beta^{\mathrm{e}}_{\mathrm{si}}  \nonumber \\
	& & + A^{f}_{\mathrm{md}}\beta^{\mathrm{e}}_{\mathrm{md}} ) z^{\mathrm{e}}_{i,n,f,\omega} + e_{(i,n)} ( y^{\mathrm{e}}_{i,n,f,\omega} + z^{\mathrm{e}}_{i,n,f,\omega} )\beta^{\mathrm{e}}_{\mathrm{ch}} \nonumber \\
	& & + \big( \frac{1}{3} ( A^{f}_{\mathrm{tx}}(\omega) \beta^{\mathrm{o}}_{\mathrm{tx}} + A^{f}_{\mathrm{rx}}(\omega)  \beta^{\mathrm{o}}_{\mathrm{rx}} ) \big) y^{\mathrm{o}}_{i,n,f,\omega}  \nonumber \\
	& & +  ( A^{f}_{\mathrm{km}} \beta^{\mathrm{o}}_{\mathrm{km}} + A^{f}_{\mathrm{si}} \beta^{\mathrm{o}}_{\mathrm{si}} + A^{f}_{\mathrm{md}} \beta^{\mathrm{o}}_{\mathrm{md}} ) z^{\mathrm{o}}_{i,n,f,\omega} \nonumber \\
	& & + e_{(i,n)} ( y^{\mathrm{o}}_{i,n,f,\omega} + z^{\mathrm{o}}_{i,n,f,\omega} ) \beta^{\mathrm{o}}_{\mathrm{ch}}  \Big)  \label{def:sp_obj} \\
\mbox{s.t.} 
	& & \sum_{ j' \in {\mathcal{O}}_{S_f} }	x_{S_f,j',f}	-	\sum_{ i' \in {\mathcal{I}}_{S_f} }	x_{i',S_f,f} =	1,	 f	\in {\mathcal{F}}	\label{eq:def_rt_const1} \\
	& & \sum_{ i' \in {\mathcal{I}}_{D_f} } x_{i', D_f, f} - \sum_{ j' \in {\mathcal{O}}_{D_f} } x_{D_f, j', f } =	1,	 f \in {\mathcal{F}}	 \label{eq:def_rt_const2} \\
	& & \sum_{ j' \in {\mathcal{O}}_n } x_{n, j', f }	-	\sum_{i' \in {\mathcal{I}}_n } x_{i',n, f}	=	0,	 f	\in {\mathcal{F}}, \nonumber \\
	& &  n \in {\mathcal{N}} \setminus \{ S_f, D_f \} \label{eq:def_rt_const3} \\
	& & \sum_{j' \in {\mathcal{O}}_n } x_{n, j', f } 	\leq	1,	 n \in {\mathcal{N}}, f \in {\mathcal{F}} \label{eq:def_rt_const4} \\ 
	& & \sum_{f \in {\mathcal{F}}} \big( y^{\mathrm{e}}_{i,j, f, \omega}  x_{i,j, f}  \big)	\leq W^{\mathrm{qkd}}_{i,j}, i,j \in {\mathcal{N}}, \forall \omega \in \Omega_{f} \label{eq:def_rt_const5} \\
	& & \sum_{f \in {\mathcal{F}}} ( z^{\mathrm{e}}_{i,j, f, \omega}  x_{i,j, f} )	\leq	W^{\mathrm{kml}}_{i,j}, i,j \in {\mathcal{N}}, \forall \omega \in \Omega_{f} 	\label{eq:def_rt_const6} \\
	& & y^{\mathrm{e}}_{i,n,f,\omega} x_{i,j, f} \leq y^{\mathrm{r}}_{i,n,f} x_{i,j,f}, i,j,n \in {\mathcal{N}}, \nonumber \\  
	& & f \in {\mathcal{F}}, \forall \omega \in \Omega_{f} \label{eq:def_rt_const7} \\
	& & z^{\mathrm{e}}_{i,n,f,\omega} x_{i,j, f} \leq z^{\mathrm{r}}_{i,n,f} x_{i,j,f}, i,j,n \in {\mathcal{N}}, \nonumber \\
	& &  f  \in {\mathcal{F}}, \forall \omega \in \Omega_{f} \label{eq:def_rt_const8}\\ 
	& & ( y^{\mathrm{e}}_{i,n,f,\omega} x_{i,j, f}) + y^{\mathrm{o}}_{i,n,f,\omega} \geq \nonumber \\ 
	& & \sum_{\omega \in \Omega_{f}} ( P_{f}({\omega}) W^{\mathrm{qkd}}_{f} x_{i, n, f} ), i,n \in {\mathcal{N}}, f	\in {\mathcal{F}} \label{eq:def_rt_const9}\\  
	& & (z^{\mathrm{e}}_{i,n,f,\omega} x_{i,j, f})  + z^{\mathrm{o}}_{i,n,f,\omega} \geq \nonumber \\  
	& & \sum_{\omega \in \Omega_{f}} ( P_{f}({\omega}) W^{\mathrm{kml}}_{f}  x_{i,n, f} ), i,n \in {\mathcal{N}}, f	\in {\mathcal{F}} \label{eq:def_rt_const10} \\
	& & x_{i,n,f} \in \{ 0, 1 \}, i,n \in {\mathcal{N}}, f \in {\mathcal{F}}  \label{eq:def_rt_const13} \\
	& & y^{\mathrm{r}}_{i, n, f }, y^{\mathrm{e}}_{i, n, f, \omega }, y^{\mathrm{o}}_{i, n, f, \omega }, z^{\mathrm{r}}_{i, n, f }, z^{\mathrm{e}}_{i, n, f, \omega }, z^{\mathrm{o}}_{i, n, f, \omega},  \nonumber \\
	& & \in \{0,1,2,\dots\}, i,n \in {\mathcal{N}}, f \in {\mathcal{F}}, \forall \omega \in \Omega_{f}  \label{eq:def_rt_const14}  	
\eeqn
\section{Performance Evaluation}
\label{sec:performance-evaluation}

In this section, we perform experiments with the following considerations. First, we consider that the proposed SP model can produce the efficient routing to satisfy the QKD-FL chain requests. Second, the deployment cost can be decreased by optimally reserving the QKD and KM wavelengths in advance through the solution of the SP model. Finally, we compare the SP model with the CO-QBN algorithm \cite{cao2021hybrid} to show the benefits of the SP model. 

\subsection{Parameter Setting}
\label{subsec:parameter-setting}

We consider the FL over QKD networks as shown in Fig. \ref{fig:system-model}. Specifically, we perform experiments on the USNET topology \cite{cao2021hybrid}. The distance $D$ between two QTxs is set to 160 km \cite{cao2021hybrid}. We initially set $W^{\mathrm{qkd}}_{i,j} = 150$ and $W^{\mathrm{kml}}_{i,j} = 50$ for the maximum wavelengths for the QKD link between node $i$ and node $j$ and the maximum wavelengths for the KM link between node $i$ and node $j$, respectively. We implement and solve the SP formulation by using GAMS/CPLEX solver \cite{Gams}. For the SP model, we consider the random number of secret-key rates with uniform distribution for ease of presentation. We consider the cost values of five QKD network components which are composed of MDI-QTXs, MDI-QRXs, LKMs, SIs, MUX/DEMUX components, and QKD and KM links. For the reservation phase, we apply the cost values from \cite{cao2021hybrid}. The cost values of the components are presented in Table \ref{table:parameter-setting}.

\begin{table}[htb] \footnotesize \caption{Reservation and on-demand cost values}
\label{table:parameter-setting}
\centering
\scalebox{0.9}{\begin{tabular}{|l|l|l|l|l|l|}\hline
{\bf Notations} & {\bf Values(\$)} & {\bf Notations} & {\bf Values(\$)} & {\bf Notations} & {\bf Values(\$)}	\\ \hline
  $\beta^{\mathrm{r}}_{\mathrm{tx}}, \beta^{\mathrm{e}}_{\mathrm{tx}}$ & 1,500 & $\beta^{\mathrm{r}}_{\mathrm{rx}}, \beta^{\mathrm{e}}_{\mathrm{rx}}$ & 2,250 & $\beta^{\mathrm{r}}_{\mathrm{km}}, \beta^{\mathrm{e}}_{\mathrm{km}}$ & 1,200 \\ \hline
  $\beta^{\mathrm{r}}_{\mathrm{si}},    \beta^{\mathrm{e}}_{\mathrm{si}}$ & 150   & $\beta^{\mathrm{r}}_{\mathrm{md}}, \beta^{\mathrm{e}}_{\mathrm{md}}$ & 300   &  $\beta^{\mathrm{r}}_{\mathrm{ch}}, \beta^{\mathrm{e}}_{\mathrm{ch}}$ & 1   \\ \hline
  $\beta^{\mathrm{o}}_{\mathrm{tx}}$ & 6,000  & $\beta^{\mathrm{o}}_{\mathrm{rx}}$  & 9,000 & $\beta^{\mathrm{o}}_{\mathrm{km}}$ & 3,000   \\ \hline
  $\beta^{\mathrm{o}}_{\mathrm{si}}$ & 500    & $\beta^{\mathrm{o}}_{\mathrm{md}}$  & 900   & $\beta^{\mathrm{o}}_{\mathrm{ch}}$ & 4    \\ \hline
\end{tabular}}
\end{table}	

\subsection{Numerical Results}
\label{subsubsec:numerical-results}

\subsubsection{Routing}
\label{subsubsec:routing}

\begin{figure}[htb]
\begin{center}
$\begin{array}{c} \epsfxsize=3.0 in \epsffile{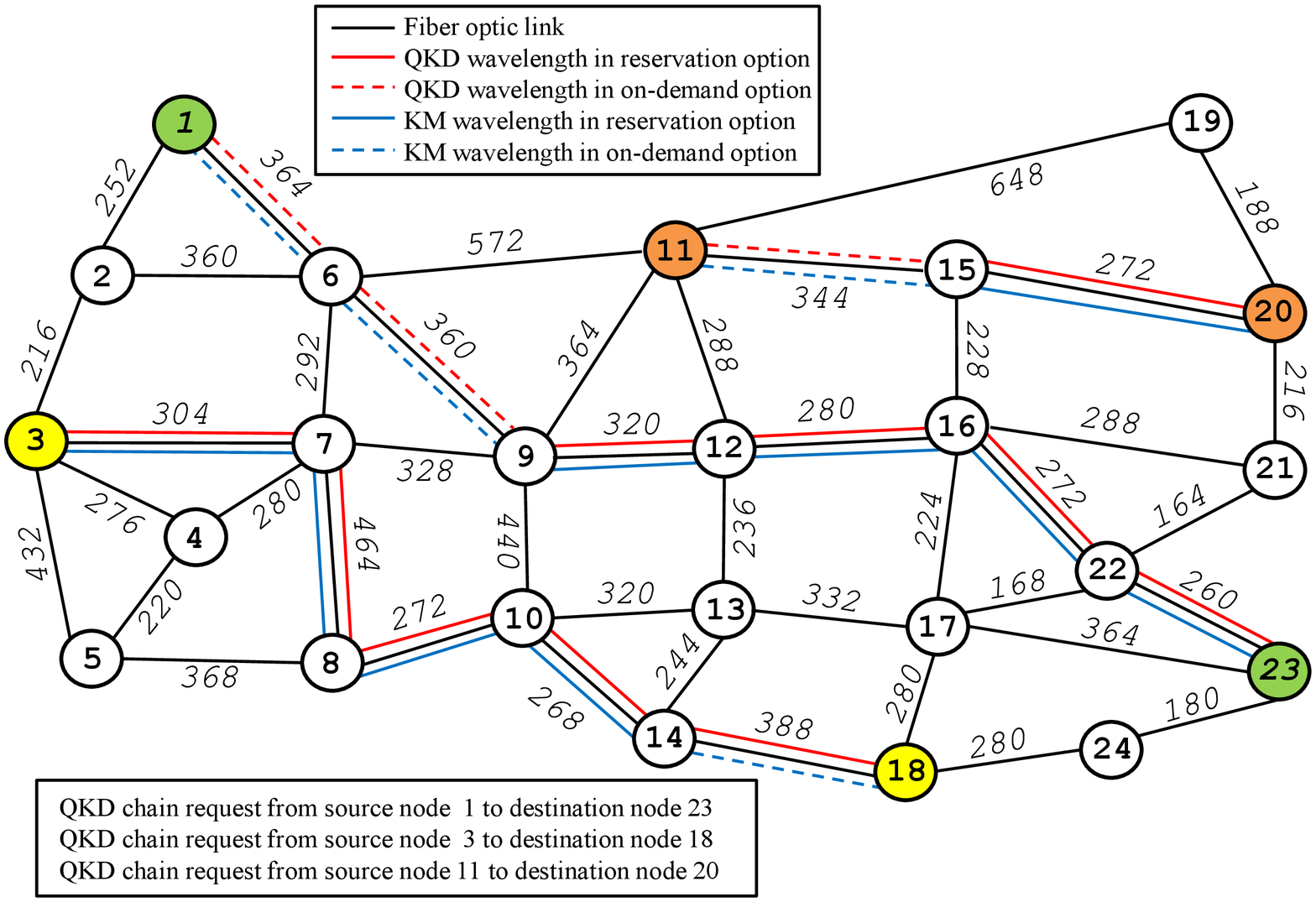} \\
\end{array}$
\caption{Three QKD-FL chain requests in USNET topology.} 
\label{fig:three-routings-fiber}
\end{center}
\vspace{-0.4cm}
\end{figure}

Figure \ref{fig:three-routings-fiber} illustrates the solutions of SP model that satisfy three requests. For the solutions, the SP model can allocate QKD and KM wavelengths (i.e., resources) in the reservation and on-demand phases for the requests in fiber optic networks. In Fig. \ref{fig:three-routings-fiber}, each request (i.e., a route) utilizes QKD and KM wavelengths within a fiber optic link. To obtain the optimal deployment cost, it is interesting to mention that both QKD and KM wavelengths in the reservation and on-demand phases are utilized along the routes. For example, with the request from source node 1 to destination node 23, the optimal deployment cost can be obtained from the route with nodes  $1 \rightarrow 6 \rightarrow 9 \rightarrow 12 \rightarrow 16 \rightarrow 22 \rightarrow 23$. In particular, the QKD and KM wavelengths in the on-demand phase are utilized from node 1 to node 6 (i.e., $1 \rightarrow 6$) and from node 6 to node 9 (i.e., $6 \rightarrow 9$) while the rest of the route utilize the QKD and KM wavelengths in the reservation phase.  

\subsubsection{Cost Structure Analysis}
\label{subsubsec:cost-stucture-analysis}

\begin{figure*}[t]
\centering
		\subfigure[The solution under different reserved QKD wavelengths.]{
			\begin{minipage}[t]{0.25\linewidth}
				\centering\includegraphics[width=1\linewidth]{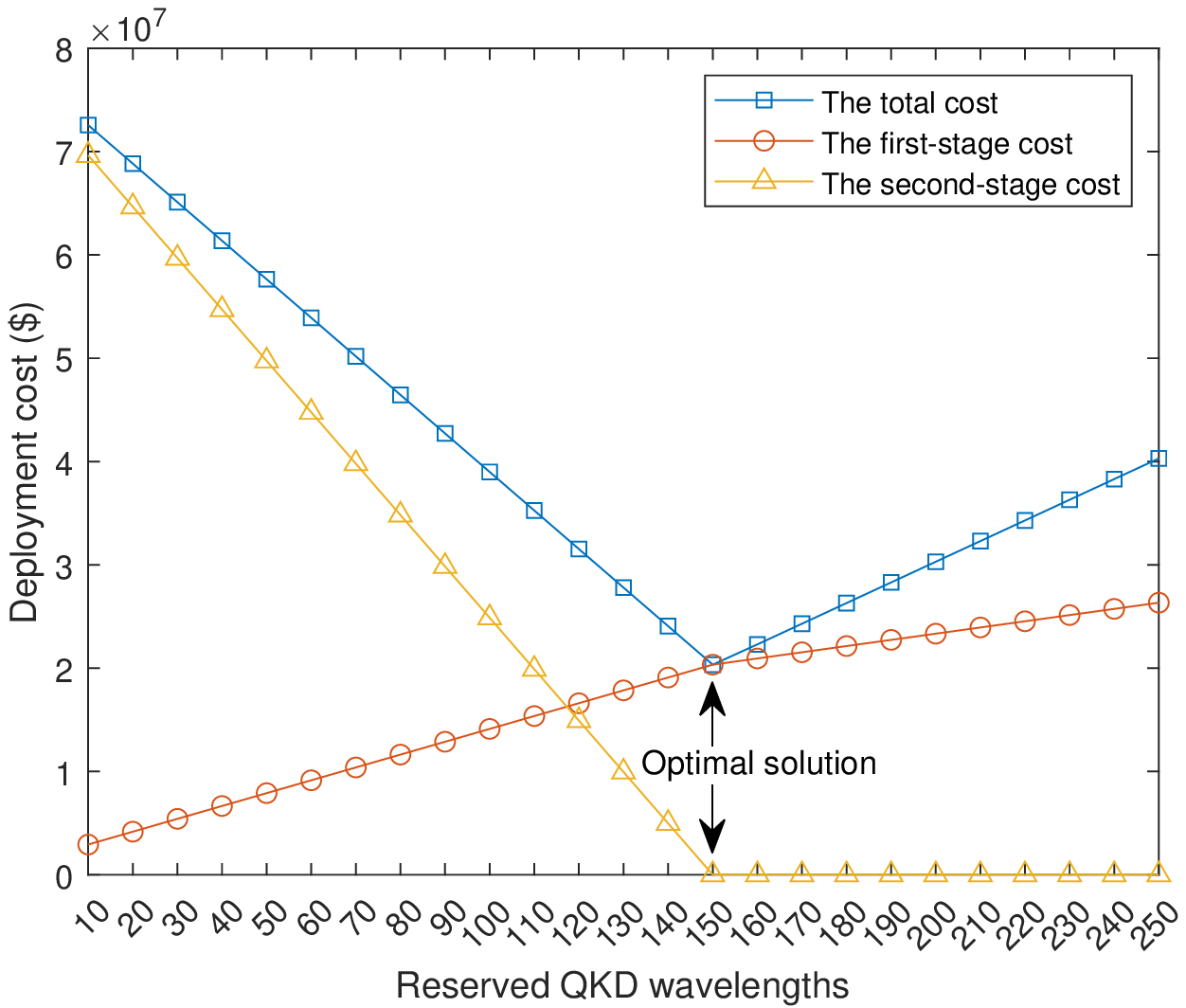}
			\end{minipage}%
		}%
		\subfigure[The solution under different reserved KM wavelengths.]{
			\begin{minipage}[t]{0.25\linewidth}
				
				\centering\includegraphics[width=1\linewidth]{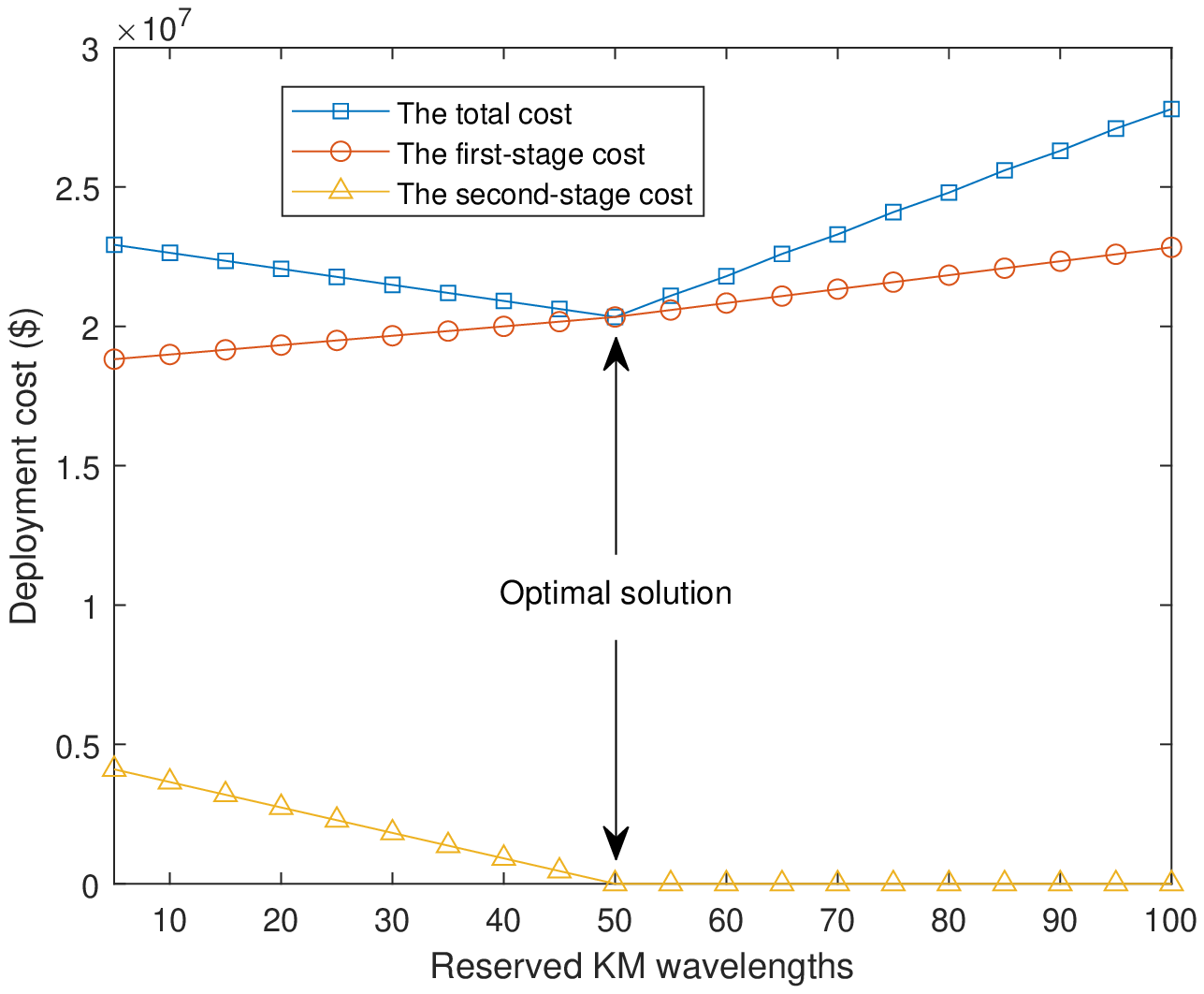}
			\end{minipage}%
		}%
		\subfigure[QKD and KM wavelength utilization.]{
			\begin{minipage}[t]{0.25\linewidth}
				
				\centering\includegraphics[width=1\linewidth]{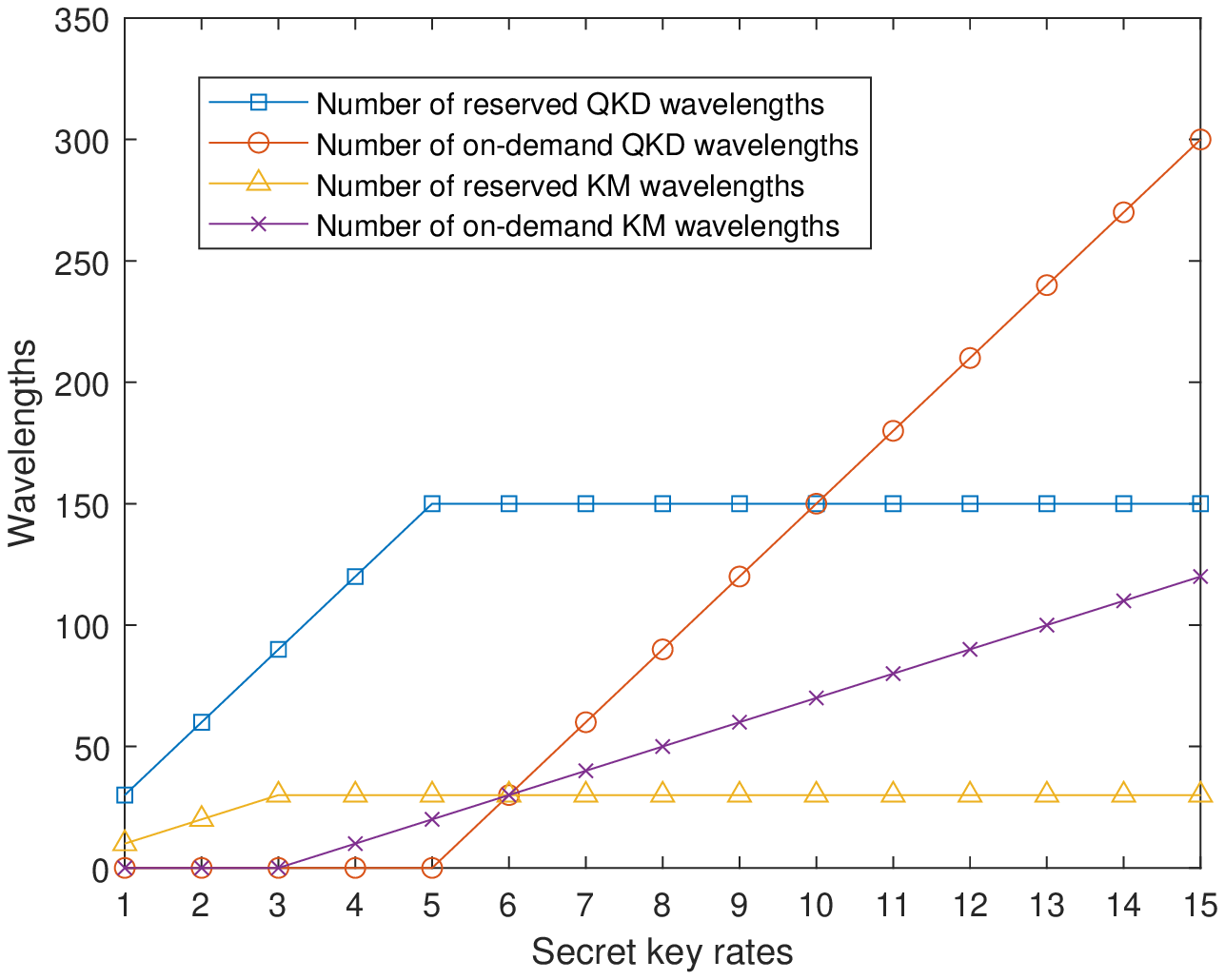}
			\end{minipage}%
		}%
		\subfigure[The deployment cost comparison.]{
			\begin{minipage}[t]{0.25\linewidth}
				
				\centering\includegraphics[width=1\linewidth]{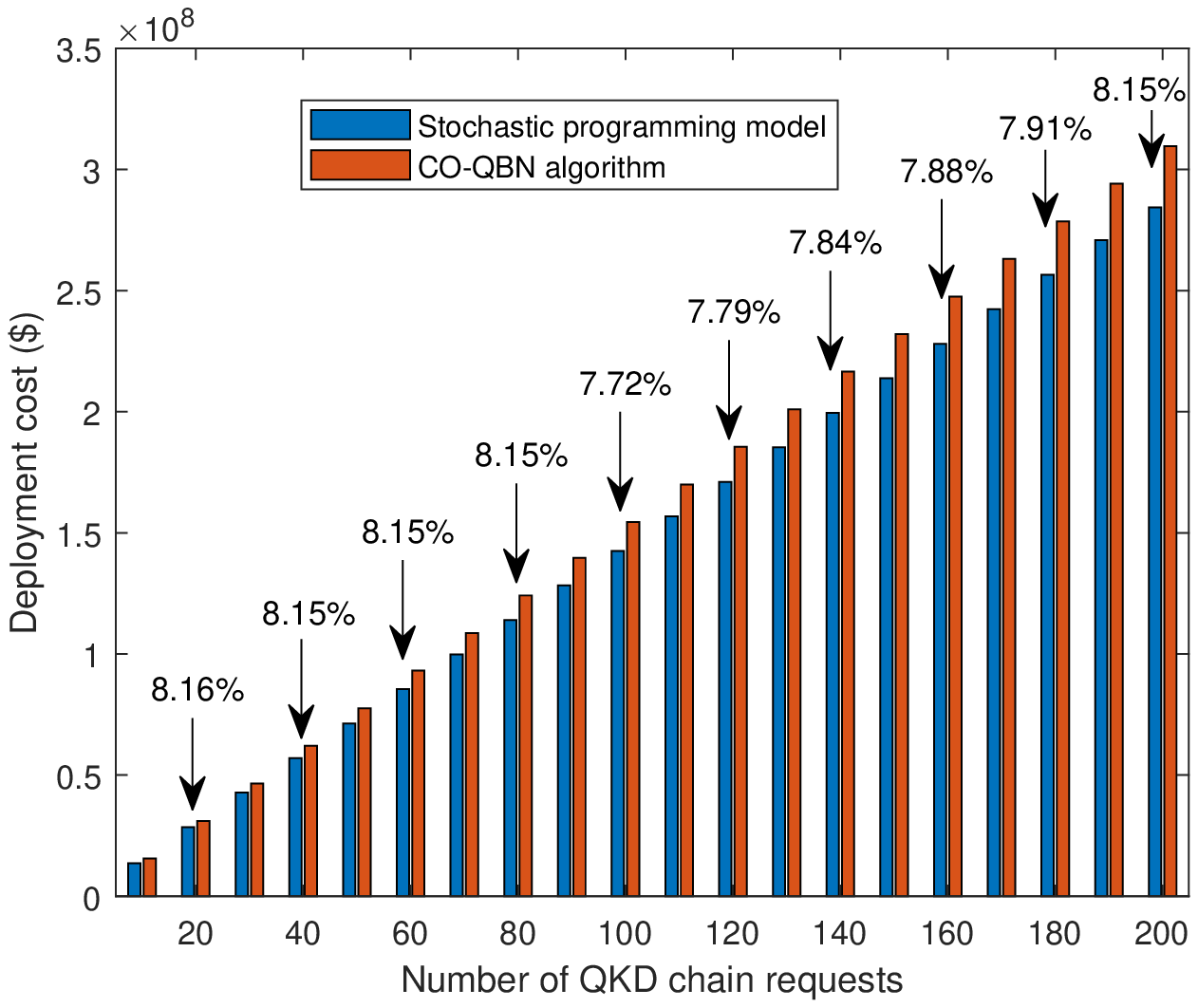}
			\end{minipage}%
		}
		\caption{(a) and (b) Optimal solutions under different reserved QKD wavelengths and different reserved KM wavelengths of the SP model, respectively, (c) The QKD and KM wavelengths in the reservation and on-demand phases under different secret-key rates, and (d) The cost comparison between the SP model and CO-QBN algorithm.}
       \label{all-figs}
       \vspace{-0.3cm}
\end{figure*}

For Fig. \ref{all-figs}(a), we examine the performance of the SP model in obtaining the optimal solution. In the first stage, we vary the reserved QKD wavelengths and fix the reserved KM wavelengths. Then, we present the optimal solution obtained by the SP model and the effect of reserved QKD wavelengths on the solution. In Fig. \ref{all-figs}(a), when the reserved QKD wavelengths increase, the first-stage cost increases significantly. However, the second-stage cost decreases dramatically when the secret-key rates are observed. The reason is that the QKD wavelength in the on-demand phase (i.e., the second stage) is forced to be minimum by utilizing the cheaper QKD wavelength in the reservation phase (i.e., the first stage). As a result, at the reserved QKD wavelengths of 150, the optimal solution can be achieved and the second-stage cost is 0. This is because the reserved QKD wavelengths meet the demands (i.e., secret-key rates) and therefore the on-demand QKD wavelengths in the second stage are not utilized. After this point, the total cost and the first-stage cost increase since there is the penalty cost for an excess of the reserved QKD wavelength to be charged. For Fig. \ref{all-figs}(a), we can mention that over-provisioning and under-provisioning of QKD wavelengths affect the high total cost significantly. 

In a similar way to the investigation above, in Fig. \ref{all-figs}(b), we vary the KM wavelengths in the first stage and fix the QKD wavelengths to show the optimal solution obtained by the SP model. In Fig. \ref{all-figs}(b), clearly, the first-stage cost increases steadily while the second-stage cost decreases. However, the optimal solution is achieved when the reserved KM wavelengths are 50. The reason is that the reserved KM wavelengths of 50 successfully satisfy the demands. After this point, the total cost and the first-stage cost increase since the penalty cost of an excess of reserved KM wavelengths is charged. Therefore, we can mention that the high cost is affected by not only the reserved QKD wavelengths but also the reserved KM wavelengths.  

Figure \ref{all-figs}(c) illustrates the QKD and KM wavelengths (i.e., resources) in the reservation and on-demand phases under different secret-key rates. In the reservation phase, the QKD and KM wavelengths rise steeply until the secret-key rates are 5 and 3 kbps, respectively. This is because the reserved QKD and KM wavelengths are completely utilized at the secret-key rates of 5 and 3 kbps, respectively. After these points, both QKD and KM wavelengths are constant since the maximum available QKD and KM wavelengths are limited. The maximum available QKD and KM wavelengths in the reservation phases are constrained by the QKD service provider. As a result, to satisfy the high secret-key rates, the QKD and KM wavelengths in the on-demand phases are instead utilized. In the on-demand phase, the QKD and KM wavelengths start rising dramatically at 5 and 3 kbps, respectively due to the limited available QKD and KM wavelengths in the reservation phases. 

\subsubsection{Performance Evaluation Under Various Parameters}
\label{subsubsec:performance-ev-various-parameters}
We compare the SP model with the CO-QBN algorithm \cite{cao2021hybrid}. Figure \ref{all-figs}(d) illustrates a performance comparison of the SP model and CO-QBN algorithm by varying the number of QKD-FL chain requests. In Fig. \ref{all-figs}(d), both the SP model and CO-QBN algorithm can decrease the deployment cost when the QKD chain requests increase. However, in comparison with the CO-QBN algorithm, the SP model can significantly achieve a lower deployment cost in different QKD chain requests. For example, with the requests of 60, the SP model can reduce the deployment cost compared with the CO-QBN algorithm by 8.15\%. In addition, the performance of the SP model compared with the CO-QBN algorithm in different requests is relatively stable (i.e., 8.21\%). Therefore, we can claim that the SP model significantly outperforms the CO-QBN algorithm.   

\section{Conclusion}
\label{sec:conclusion}
In this paper, we have proposed the architecture and adaptive QKD resource allocation and routing model for QKD-FL systems. First, we have proposed the hierarchical architecture for QKD-FL systems for the joint allocation of resources for QKD nodes and links to protect the transmission of FL models from eavesdropping attacks. Then, we have formulated the adaptive QKD resource allocation and routing model based on two-stage SP to obtain the optimal deployment cost under uncertainty. The experimental results have clearly shown that the proposed model produces the minimum deployment cost and shortest routing. In addition, the performance of the proposed model outperforms that of the CO-QBN algorithm by at least 7.72\%. 

For the future work, we will investigate the impact of the uncertainty (i.e., secret-key rates). In addition, we will study and propose a resource allocation and routing model in QKD over space-air-ground integrated networks (SAGIN) for FL applications.


	\bibliographystyle{ieeetr}
	\bibliography{networkf-conf}
\end{document}